\documentclass[english]{IEEEtran}
\usepackage[T1]{fontenc}
\usepackage[latin9]{inputenc}
\usepackage{array}
\usepackage{verbatim}
\usepackage{rotfloat}
\usepackage{booktabs}
\usepackage{amsmath}
\usepackage{graphicx}
\usepackage{esint}

\makeatletter

\providecommand{\tabularnewline}{\\}

\makeatother

\usepackage{babel}
\begin{document}

\title{Fourier, Gabor, Morlet or Wigner:\\
Comparison of Time-Frequency Transforms}

\author{Stefan Scholl\\
dc9st@panoradio-sdr.de}
\maketitle
\begin{abstract}
In digital signal processing time-frequency transforms are used to
analyze time-varying signals with respect to their spectral contents
over time. Apart from the commonly used short-time Fourier transform,
other methods exist in literature, such as the Wavelet, Stockwell
or Wigner-Ville transform. Consequently, engineers working on digital
signal processing tasks are often faced with the question which transform
is appropriate for a specific application. To address this question,
this paper first briefly introduces the different transforms. Then
it compares them with respect to the achievable resolution in time
and frequency and possible artifacts. Finally, the paper contains
a gallery of time-frequency representations of numerous signals from
different fields of applications to allow for visual comparison.
\end{abstract}

\section{Introduction}

In many fields of engineering and science it is vital to analyze the
spectral content of time-varying signals as it changes over time.
This includes medical signals (EEG, ECG, ultrasonic), music and speech,
animal voices, seismic activity, vibrations, fluctuations in power
grids, radar and communication signals and many more.

For that purpose many different types of time-frequency transforms
have been introduced in literature. Some of the transforms have a
high practical impact, others are more of theoretical interest. The
available methods provide largely different properties such as time
resolution, frequency resolution, accuracy and the potential introduction
of artifacts that do not correspondent to actual signal components.

This paper considers the STFT, Gabor, Wavelet and Stockwell (S) transform
as well as the Wigner-Ville distribution (WVD) and its smoothed pseudo
variant (SPWVD). Section \ref{sec:Time-Frequency-Transforms} provides
a brief introduction to these methods. Readers that are interested
in more in-depth information and the theoretical backgrounds are referred
to \cite{semmlow2014biosignal,sandsten2018,Hlawatsch1992}. Section
\ref{sec:Comparison} compares the transforms with respect to resolution
and artifacts. Finally, the transforms are applied to numerous different
signals to form a gallery of time-frequency transforms. The purpose
of this gallery is to show the properties, advantages and disadvantages
of the transforms on actual signals in order to support engineers
selecting the right transform for their application.

\section{\label{sec:Time-Frequency-Transforms}Time-Frequency Transforms}

\subsection{Short-Time Fourier and Gabor Transform}

The STFT is the most widely known and commonly used time-frequency
transform. It is well understood, easy to interpret and there exist
fast implementations (FFT). Its drawbacks are the limited and fixed
resolution in time and frequency.

The idea of the STFT is to move a sliding window $w(t)$ over the
signal $x(t)$ to be analyzed, such that a particular time span of
the signal is selected. For each position of the window a Fourier
transform is calculated, that represents the frequency content of
that time span. The STFT results in the two dimensional time-frequency
representation:

\[
X(t,f)=\intop_{-\infty}^{\infty}x(t_{1})w^{*}(t_{1}-t)e^{-j2\pi ft_{1}}dt_{1}
\]

For spectral analysis the squared magnitude of the STFT is considered,
which is called spectrogram:

\[
S_{x}(t,f)=\left|X(t,f)\right|^{2}
\]

\begin{figure}
\begin{centering}
\includegraphics[bb=0bp 0bp 527bp 300bp,clip,scale=0.3]{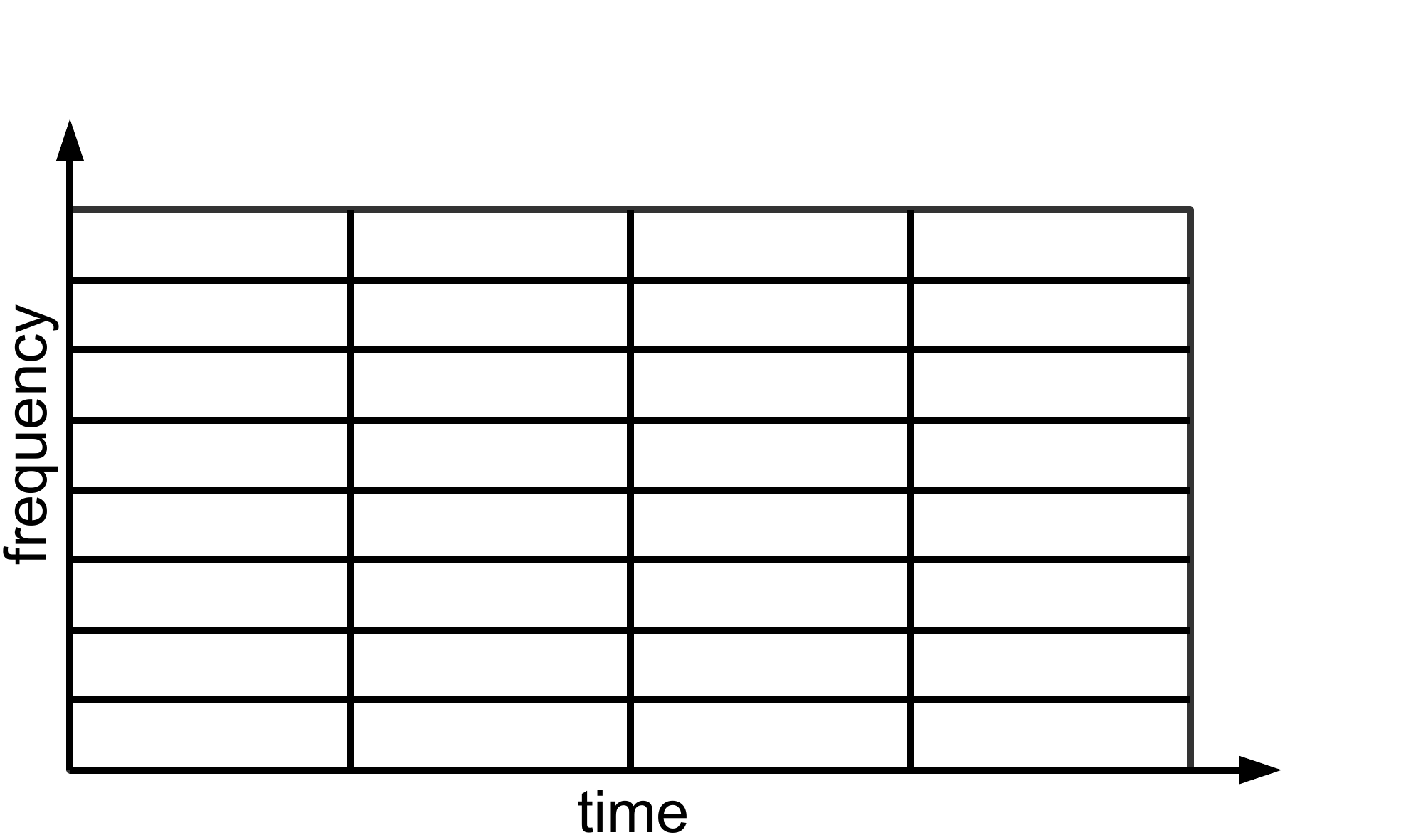}
\par\end{centering}

\begin{centering}
\includegraphics[bb=0bp 0bp 527bp 300bp,clip,scale=0.3]{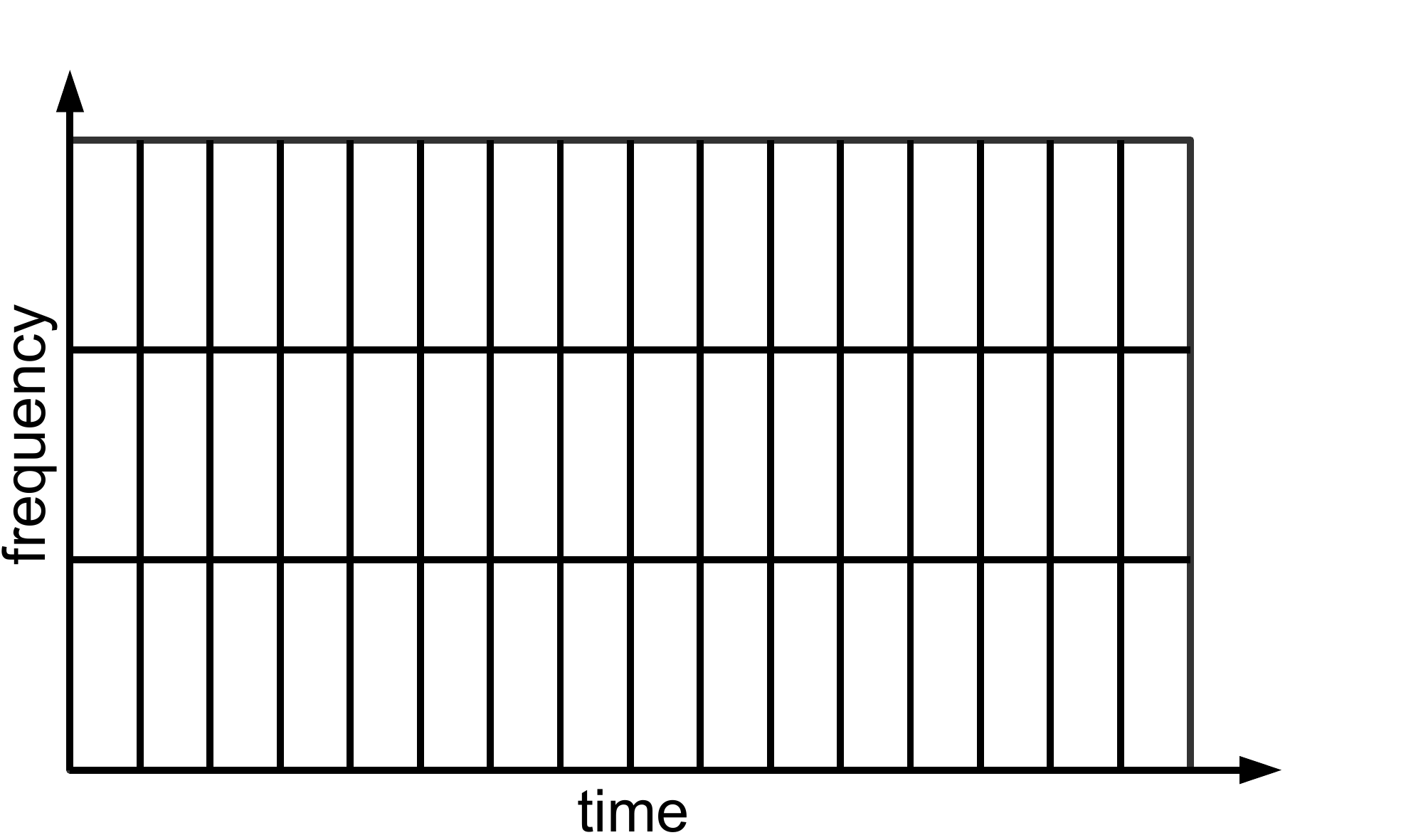}
\par\end{centering}

\caption{\label{fig:stft_resolution}Resolution of the STFT with long (upper)
and short window (lower)}
\end{figure}

Time and frequency resolution can be controlled by the window length,
as shown in Figures \ref{fig:STFT} and \ref{fig:stft_resolution}:
A short window captures only a short period of time and has thus a
precise time resolution. However, the frequency resolution is poor,
because the windowed signal contains only few time samples resulting
in only few frequency bins. Contrary, a long window provides poor
time resolution, but creates precise frequency information due to
the larger number of samples. This phenomenon is known as the uncertainty
principle, that states that the product of resolution in time and
frequency is limited: $BT\geq\frac{1}{4\pi}$ (with B being the bandwidth
of one frequency bin) \cite{sandsten2018}. 

A special case of the STFT, where the uncertainty equation above is
fulfilled with equality (i.e. having the best joint time and frequency
resolution), is known as the Gabor transform. The Gabor transform
is simply a STFT with the window being a Gaussian function

\[
w(t)=e^{-\alpha t^{2}}
\]

where the parameter $\alpha$ controls the window length, i.e. the
emphasis on time or frequency resolution. 

\begin{figure}
\begin{centering}
\includegraphics[scale=0.5]{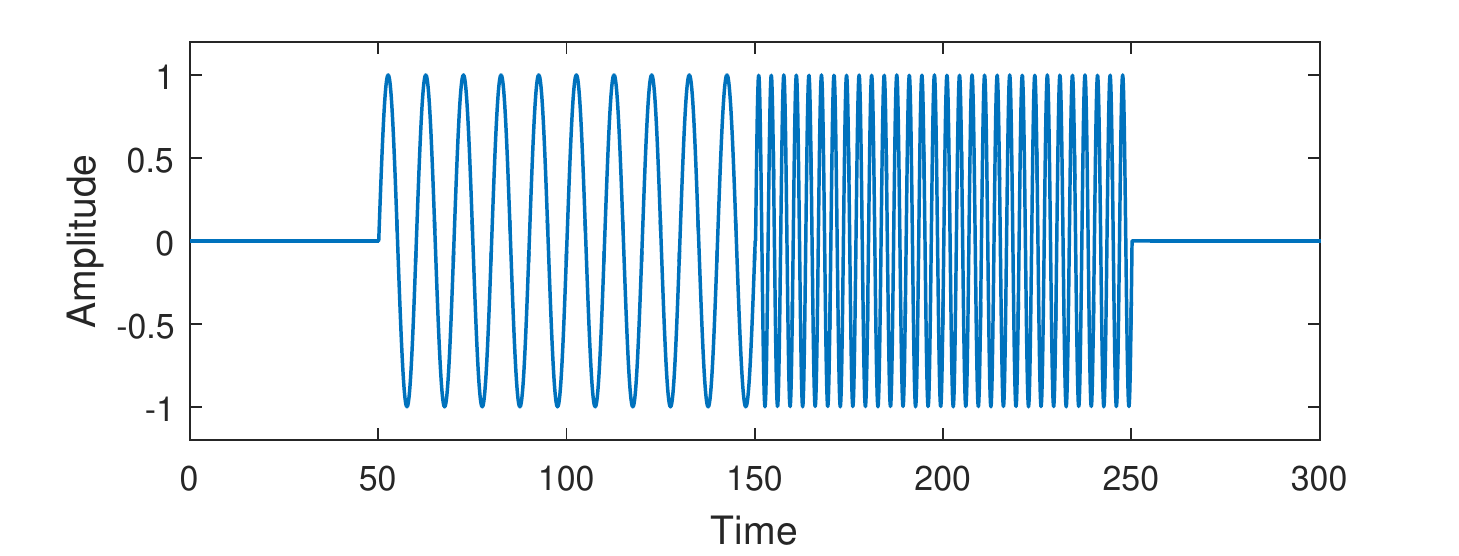}
\par\end{centering}

\caption{\label{fig:Exemplary-signal-in}Exemplary signal in the time domain}

\end{figure}

\begin{figure}
\begin{centering}
\includegraphics[bb=30bp 30bp 410bp 440bp,clip,scale=0.5]{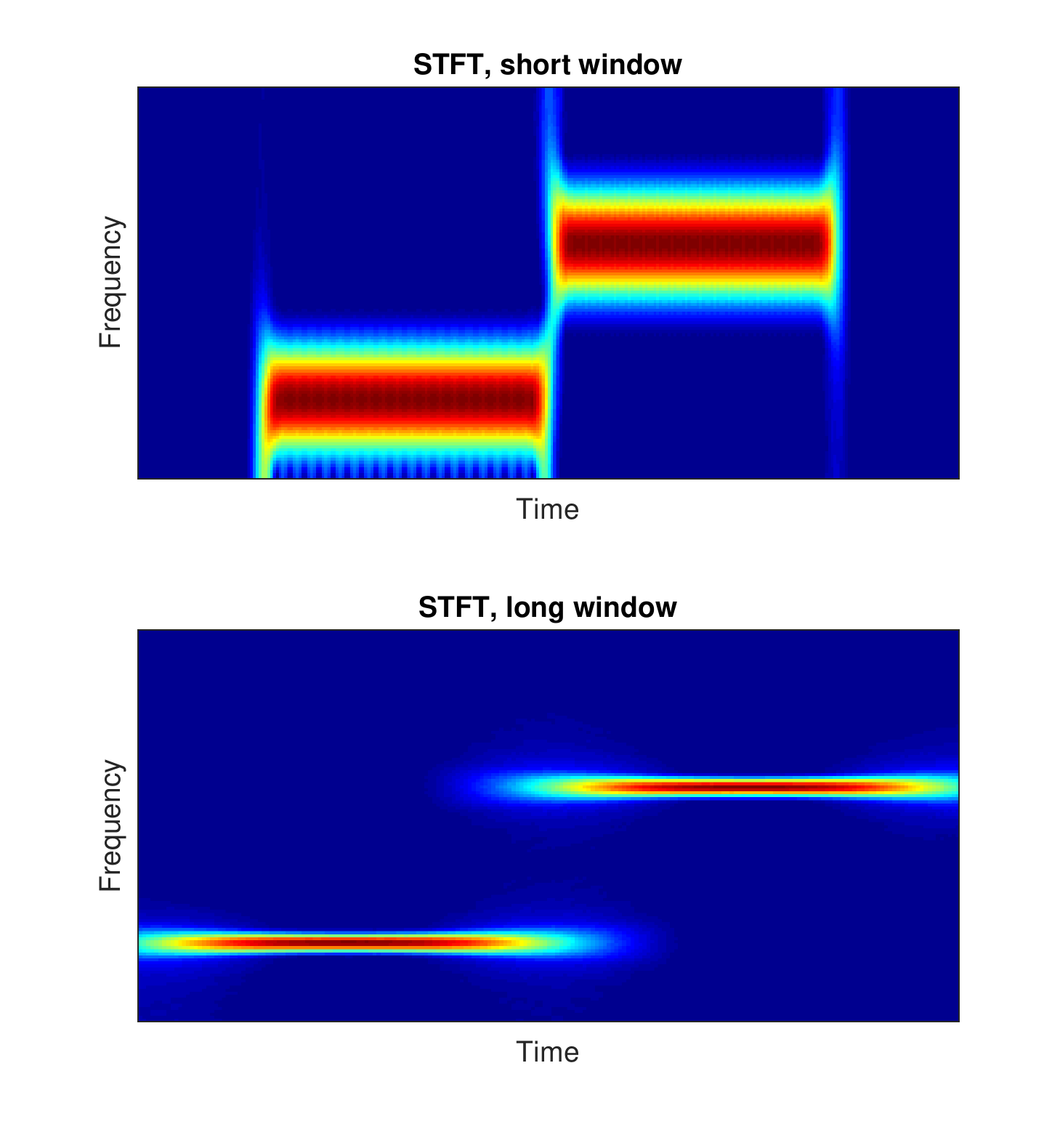}
\par\end{centering}

\caption{\label{fig:STFT}STFT (spectrogram) with short and long windows of
the exemplary signal in Figure \ref{fig:Exemplary-signal-in}, showing
the difference in frequency and time resolution}
\end{figure}

\subsection{Wavelet Transform}

\begin{figure}
\begin{centering}
\includegraphics[bb=0bp 0bp 527bp 300bp,clip,scale=0.3]{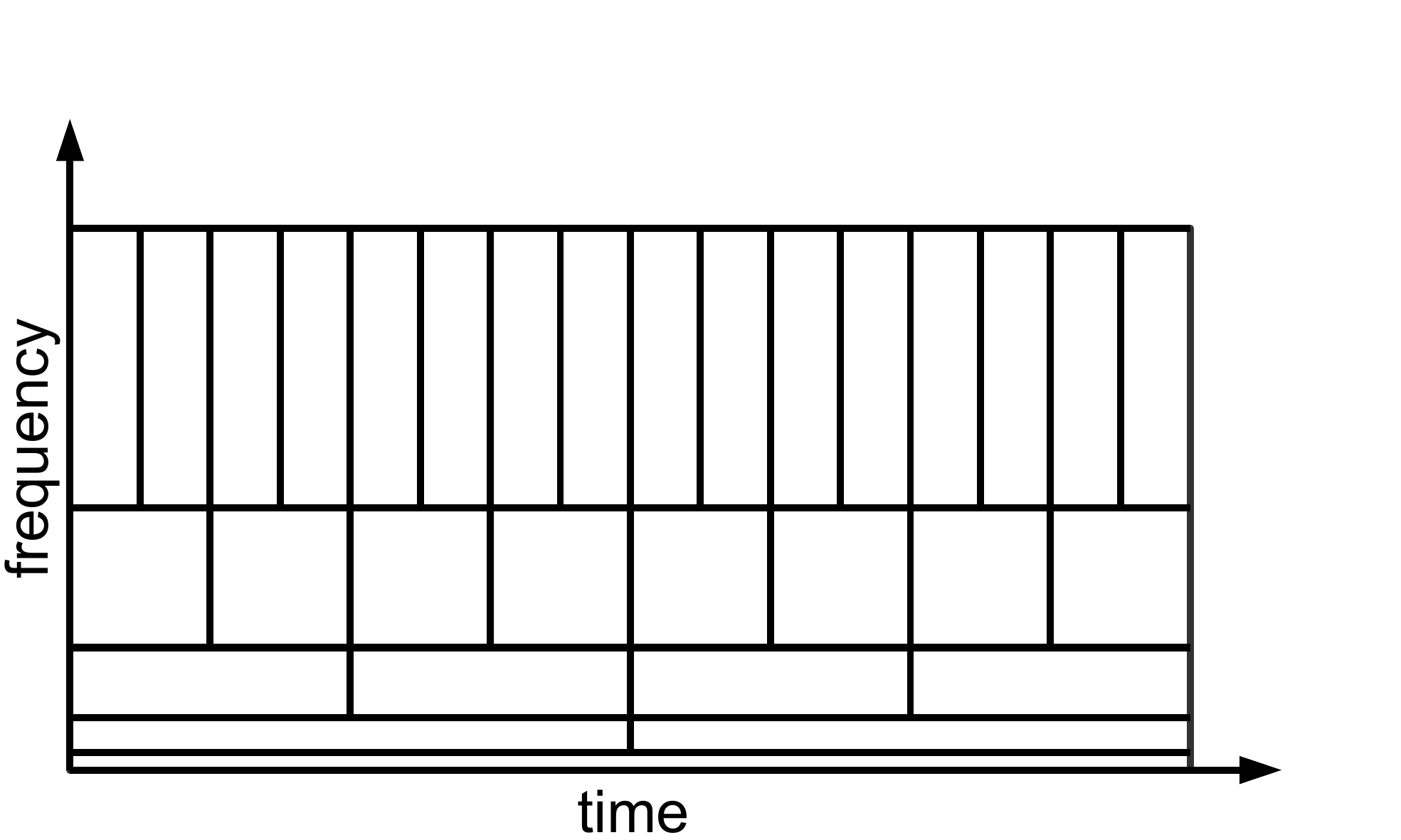}
\par\end{centering}

\caption{\label{fig:Variable-resolution-property-cwt}Variable resolution property
of the CWT and S transform}
\end{figure}

The result of the wavelet transform differs from the STFT in that
its time-frequency resolution is not fixed and depends on the frequency
(multi-scale property, see Fig. \ref{fig:Variable-resolution-property-cwt}).
In general, the wavelet transform represents lower frequency components
with finer frequency resolution and coarser time resolution. For higher
frequencies the reverse is true: frequency resolution is coarser and
time resolution is finer. This variable resolution property of the
wavelet transform is sometimes superior to the Fourier approach, because
it may give clearer spectral information for certain applications,
such as audio signal processing.

The wavelet transform compares the time domain signal $x(t)$ with
a short analysis function $\varPsi(t)$. $\varPsi(t)$ is called the
wavelet and can take on many forms as will be described below. During
the calculation of the transform the wavelet is repeatedly moved over
the signal (time shifted) -- each pass with a different scale factor
in time, that dilates the wavelets to a different length (dilation
or scale). This creates a two dimensional representation of time (i.e.
shift) and scale (can be related to frequency). The time shift is
denoted by $b$, the scale by $a$. The continuous wavelet transform
(CWT) is defined as:

\[
CWT(a,b)=\frac{1}{\sqrt{a}}\intop_{-\infty}^{\infty}x(t)\varPsi^{*}\left(\frac{t-b}{a}\right)dt
\]

Analog to the spectrogram of the STFT, the scalogram of the wavelet
transform is defined as

\[
\left|CWT(a,b)\right|^{2}
\]

Different wavelet functions are available for analysis. For the CWT
commonly Morlet wavelets (also called Gabor wavelets) are used, that
consist of a complex sine wave with Gaussian envelope 
\[
\varPsi_{Morlet}(t)=e^{-\alpha t^{2}}e^{j2\pi f_{c}t}
\]

\begin{figure}
\begin{centering}
\includegraphics[bb=35bp 15bp 380bp 305bp,clip,scale=0.5]{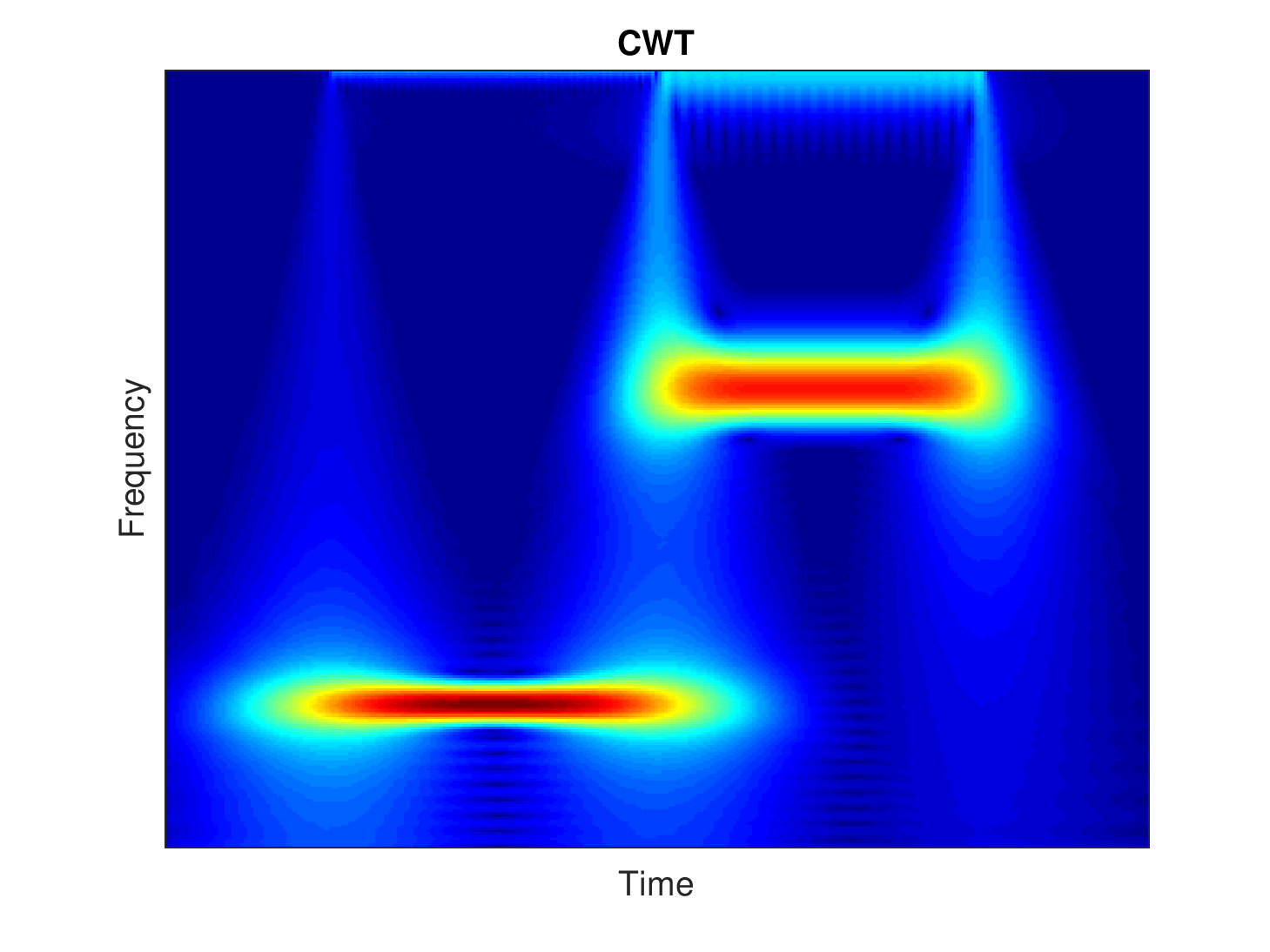}
\par\end{centering}

\caption{\label{fig:CWT}CWT (scalogram): note the varying resolution of frequency
and time}
\end{figure}

Here, the parameters ``center frequency'' $f_{c}$ and ``width
of Gaussian'' $\alpha$ control the trade-off between time and frequency
resolution and need to be selected before conducting the transform.
Basically, wavelet functions need to be functions that are both local
in time and frequency to provide adequate time and frequency resolution.
Figure \ref{fig:CWT} shows an example of the scalogram of a CWT. 

Note, that besides the CWT, the discrete wavelet transform (DWT) exists.
Different from the CWT, the DWT preserves the complete information
of the time domain signal and is thus invertible. Therefore the DWT
is often used to create sparse signal representations that can be
used for data compression (e.g. widely applied in image processing).
Usually different types of wavelets are used for the DWT, such as
the Haar or Daubechies wavelets. The DWT is often considered less
suitable for time-frequency representation, because it creates less
readable plots.

\subsection{Stockwell Transform}

The Stockwell or S-transform \cite{Stockwell1996} is basically a
STFT with a Gaussian (Gabor) window, whose length is frequency dependent.
This results in a varying time-frequency resolution similar to the
wavelet transform (Figure \ref{fig:Variable-resolution-property-cwt}).
The Stockwell transform is defined as

\[
ST(t,f)=\intop_{-\infty}^{\infty}x(t_{1})\frac{\left|f\right|}{\sqrt{2\pi}}e^{\frac{-f^{2}(t_{1}-t)^{2}}{2}}e^{-j2\pi ft_{1}}dt_{1}
\]

\begin{figure}
\begin{centering}
\includegraphics[bb=35bp 15bp 380bp 305bp,clip,scale=0.5]{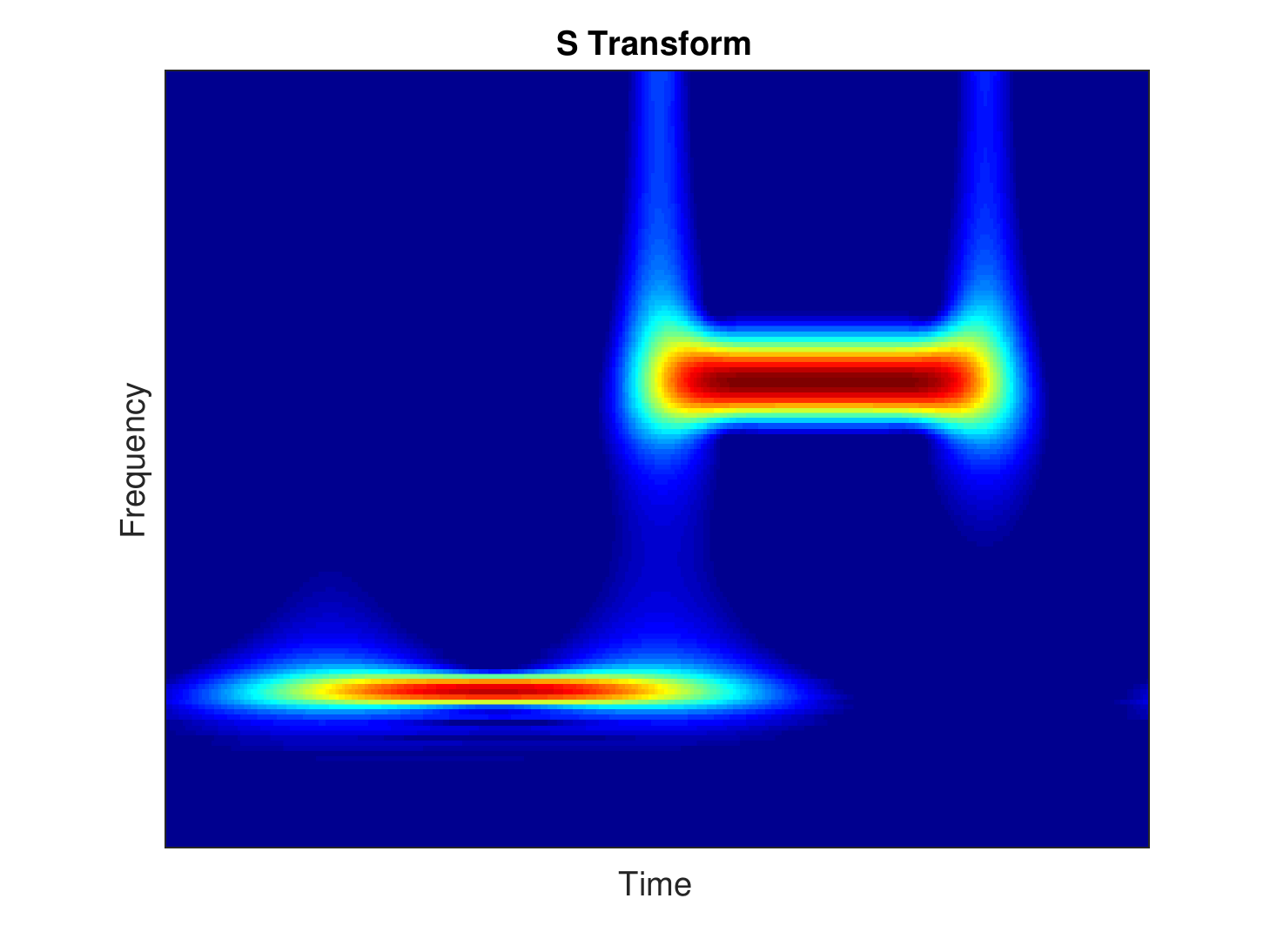}
\par\end{centering}

\caption{\label{fig:S-transform}S transform: note the varying resolution of
frequency and time and the emphasis on the higher frequency component}
\end{figure}

It is also similar to the CWT with a Morlet wavelet. However, in contrast
to the CWT, the Stockwell transform has absolute referenced phase
information. This means that the phase of the kernel functions, which
are multiplied with the signal for analysis, at t = 0 is zero. Moreover,
the Stockwell transform tends to emphasize higher frequency content
due to the factor $|f|$ in its formula. Figure \ref{fig:S-transform}
shows an example.

\subsection{Wigner-Ville Distribution}

The Wigner-Ville distribution (WVD) \cite{semmlow2014biosignal} overcomes
the limited resolution of the Fourier and wavelet based methods using
an autocorrelation approach. 

The standard version of the autocorrelation function (ACF) considers
the pointwise multiplication of a signal with a lagged version of
itself and integrates the results over time. It is defined as

\[
r{}_{xx}(\tau)=\intop_{-\infty}^{\infty}x(t)x^{*}(t+\tau)dt
\]

The standard ACF is only dependent on the lag $\tau$, because time
is integrated out of the result. The WVD uses a variation of the ACF,
called the instantaneous autocorrelation, which omits the integration
step. Thus time remains in the result. The instantaneous autocorrelation
is therefore a two dimensional function, depending on $t$ and the
lag $\tau$:

\[
R_{xx}(t,\tau)=x(t+\frac{\tau}{2})x^{*}(t-\frac{\tau}{2})
\]

\begin{figure}
\begin{centering}
\includegraphics[scale=0.55]{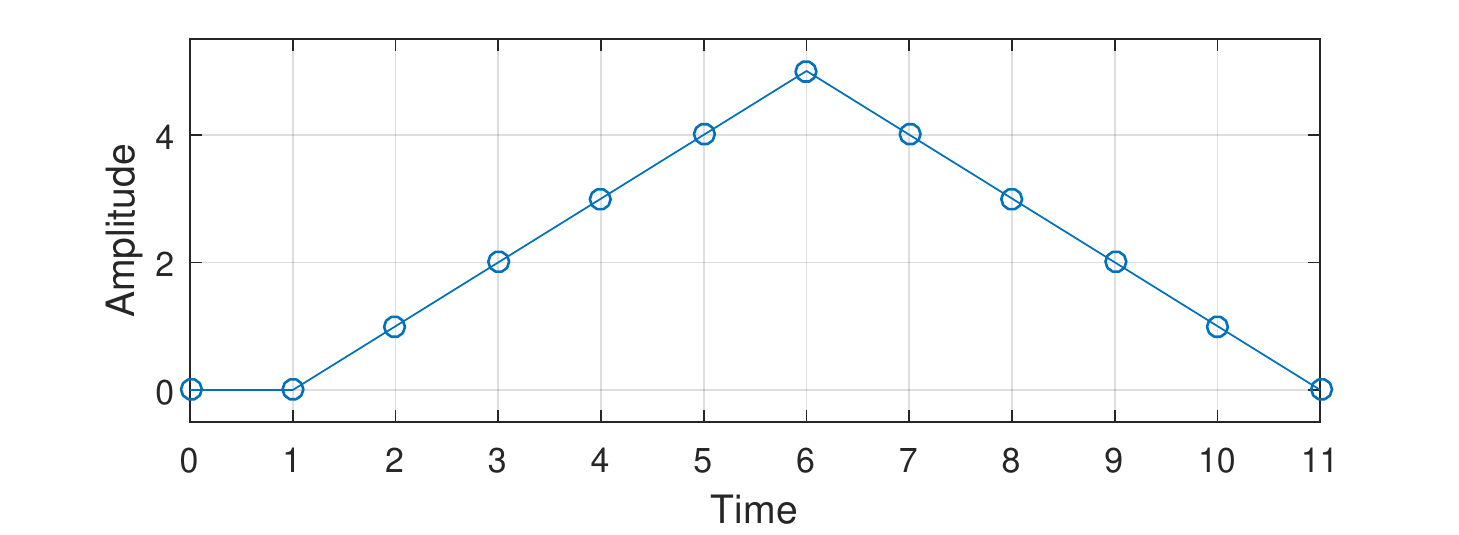}
\par\end{centering}

\begin{centering}
\includegraphics[clip,scale=0.6]{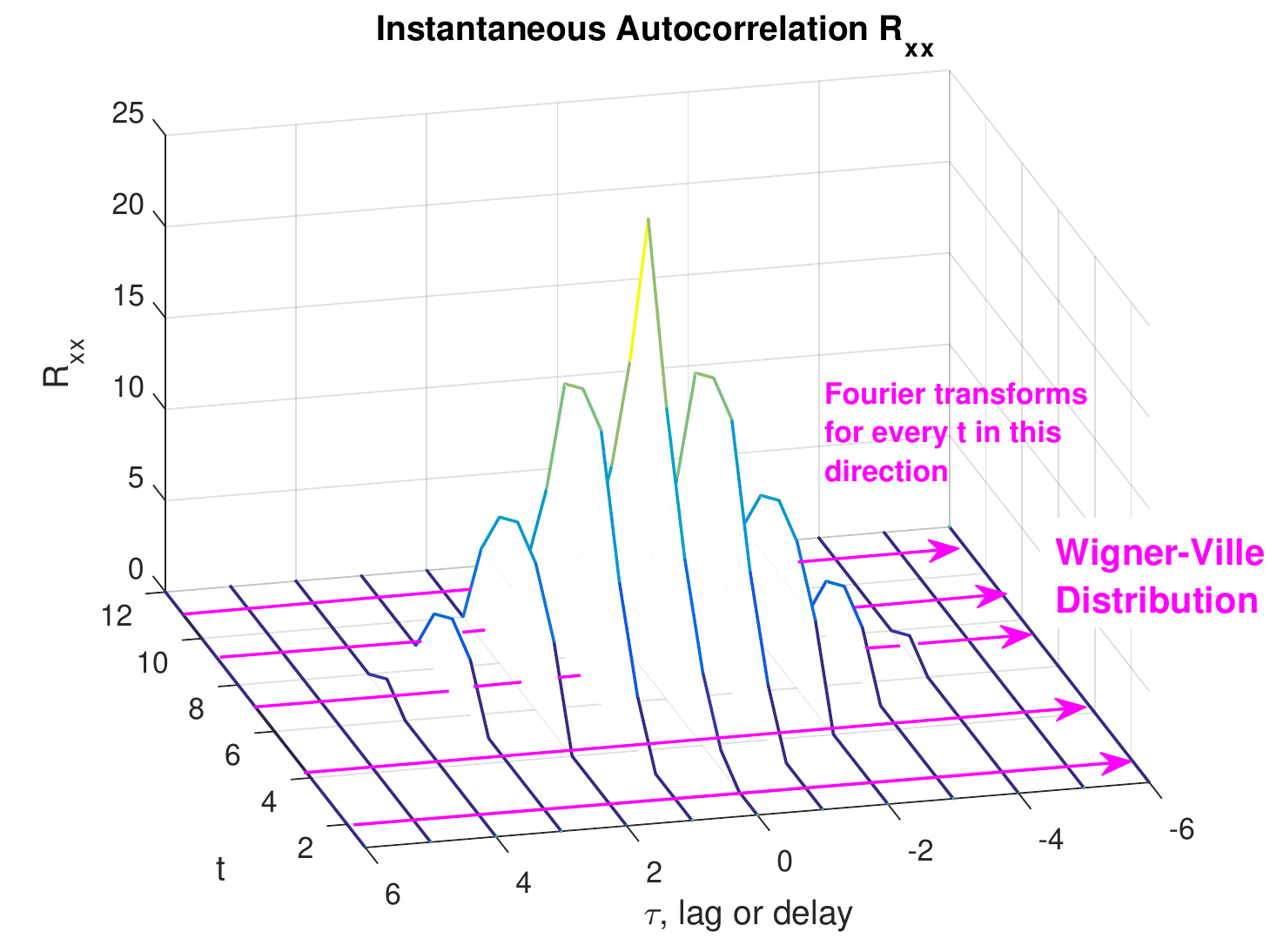}
\par\end{centering}

\caption{\label{fig:wvd-calculation}Depiction of the instantaneous autocorrelation
$R_{xx}(t,\tau)$ and the calculation of the WVD of a triangular signal.
The slices represent the pointwise multiplication of the signal with
a lagged version of itself for a specific lag value $\tau$. Calculating
a Fourier transform across the $\tau$-axis for every value of $t$
creates the WVD. If one would integrate over $t$ instead, the result
would be the standard ACF.}
\end{figure}

As an example, Figure \ref{fig:wvd-calculation} shows the instantaneous
autocorrelation of a triangular shaped signal. 

The WVD calculates the frequency content for each time step $t$ by
taking a Fourier transform of the instantaneous autocorrelation across
the axis of the lag variable $\tau$ for that given $t$ (depicted
in Figure \ref{fig:wvd-calculation}) :

\begin{eqnarray*}
W(t,f) & = & \intop_{-\infty}^{\infty}R_{xx}(t,\tau)e^{-j2\pi f\tau}d\tau\\
 & = & \intop_{-\infty}^{\infty}x(t+\frac{\tau}{2})x^{*}(t-\frac{\tau}{2})e^{-j2\pi f\tau}d\tau
\end{eqnarray*}

The result is real-valued. This way of calculation is related to the
fact, that the Fourier spectrum of a signal equals the Fourier transform
of its ACF.

The WVD offers very high resolution in both time and frequency, this
is much finer than of the STFT.

\begin{figure}
\begin{centering}
\includegraphics[bb=35bp 15bp 400bp 320bp,clip,scale=0.5]{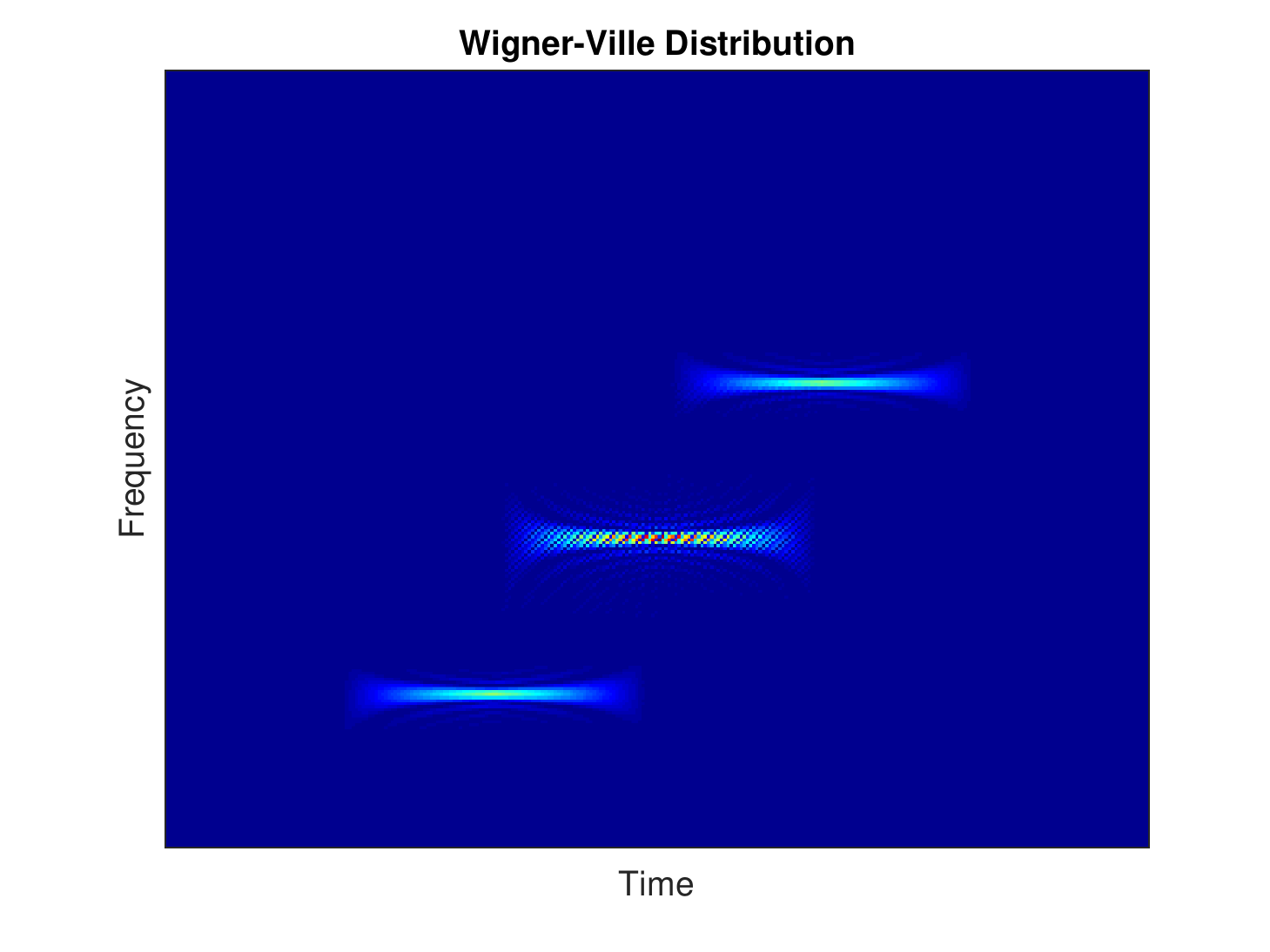}
\par\end{centering}

\caption{\label{fig:WVD}Wigner-Ville distribution: Cross term artifacts occur
with high amplitude and oscillating behavior}
\end{figure}

An important disadvantage of the WVD are the so-called cross terms.
These are artifacts occurring in the result, if the input signal contains
a mixture of several signal components, see Figure \ref{fig:WVD}.
They stem from the fact, that the WVD is a quadratic (and therefore
a non-linear) transform due to the way the instantaneous autocorrelation
is calculated. The WVD of the superposition of two signals is

\[
W{}_{x_{1}+x_{2}}=W{}_{x_{1}}+W{}_{x_{2}}+2\Re\left\{ W_{x_{1},x_{2}}\right\} 
\]

and may be dominated by the cross term $W_{x_{1},x_{2}}$, which may
have twice the amplitude of the auto terms $W{}_{x_{1}}$and $W{}_{x_{2}}$.
Unfortunately, the occurrence of these cross terms limits the usefulness
for many practical signals.

The WVD is usually calculated with the analytic version of the input
signal that does not contain negative frequency components. This avoids
cross terms between positive and negative frequency content that may
mask low frequency components in the WVD. 

The cross terms occur midway between the auto terms and often have
an oscillatory (high-frequency) pattern. A method to reduce cross
terms is to suppress the oscillating components by additional low-pass
filtering in time and frequency. However, this suppression of cross
terms comes at the expense of reduced resolution. This idea of additional
cross term suppression leads to the more general formulation of time-frequency
transforms called Cohen's class \cite{semmlow2014biosignal,Hlawatsch1992}.
From Cohen's class many different variants can be deduced, that basically
differ in the way the low-pass filter is designed. A prominent one
is the smoothed pseudo Wigner-Ville distribution (SPWVD) \cite{Hlawatsch1992}.
Other variants such as Choi-Williams, Margenau-Hill or Rihaczek can
be found in literature, but often provide very similar results to
the SPWVD for practical signals. The SPWVD is defined as the WVD filtered
by two separate kernels $g(t)$ and $H(f)$ (need to be chosen prior
to the transform), that smooth the WVD in frequency and time:

\[
SPWVD(t,f)=\intop_{t_{1}}\intop_{f_{1}}g(t-t_{1})H(f-f_{1})W(t_{1,}f_{1})dt_{1}df_{1}
\]

\begin{figure}
\begin{centering}
\includegraphics[bb=35bp 15bp 400bp 320bp,clip,scale=0.5]{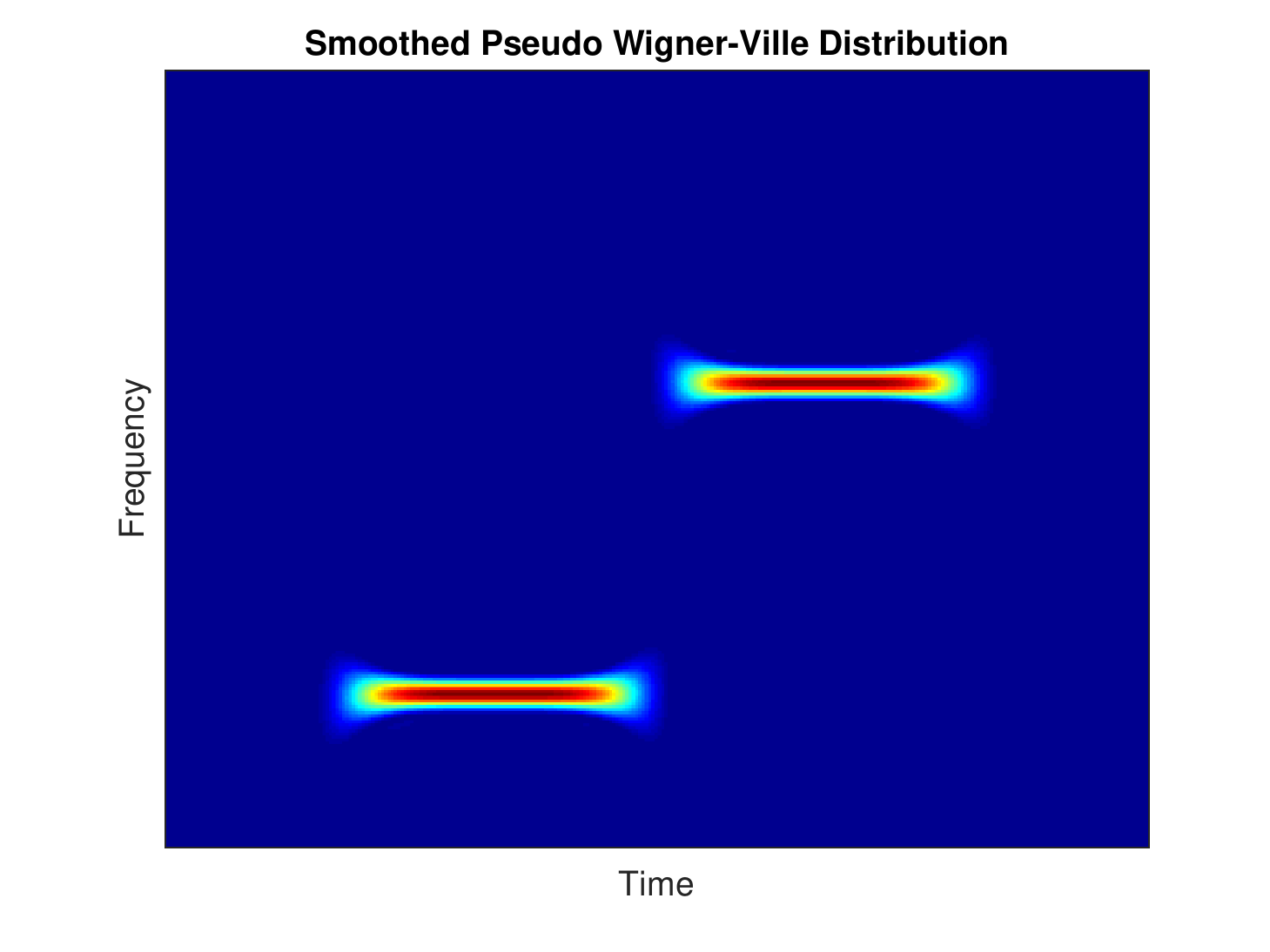}
\par\end{centering}

\caption{\label{fig:SPWVD}Smoothed pseudo WVD: cross terms are suppressed}
\end{figure}

Figure \ref{fig:WVD} shows an example of the WVD and Figure \ref{fig:SPWVD}
its smoothed version, the SPWVD.

\section{\label{sec:Comparison}Comparison and Time-Frequency Gallery}

\begin{figure*}
\begin{centering}
\includegraphics[bb=10bp 150bp 1111bp 1644bp,clip,scale=0.3]{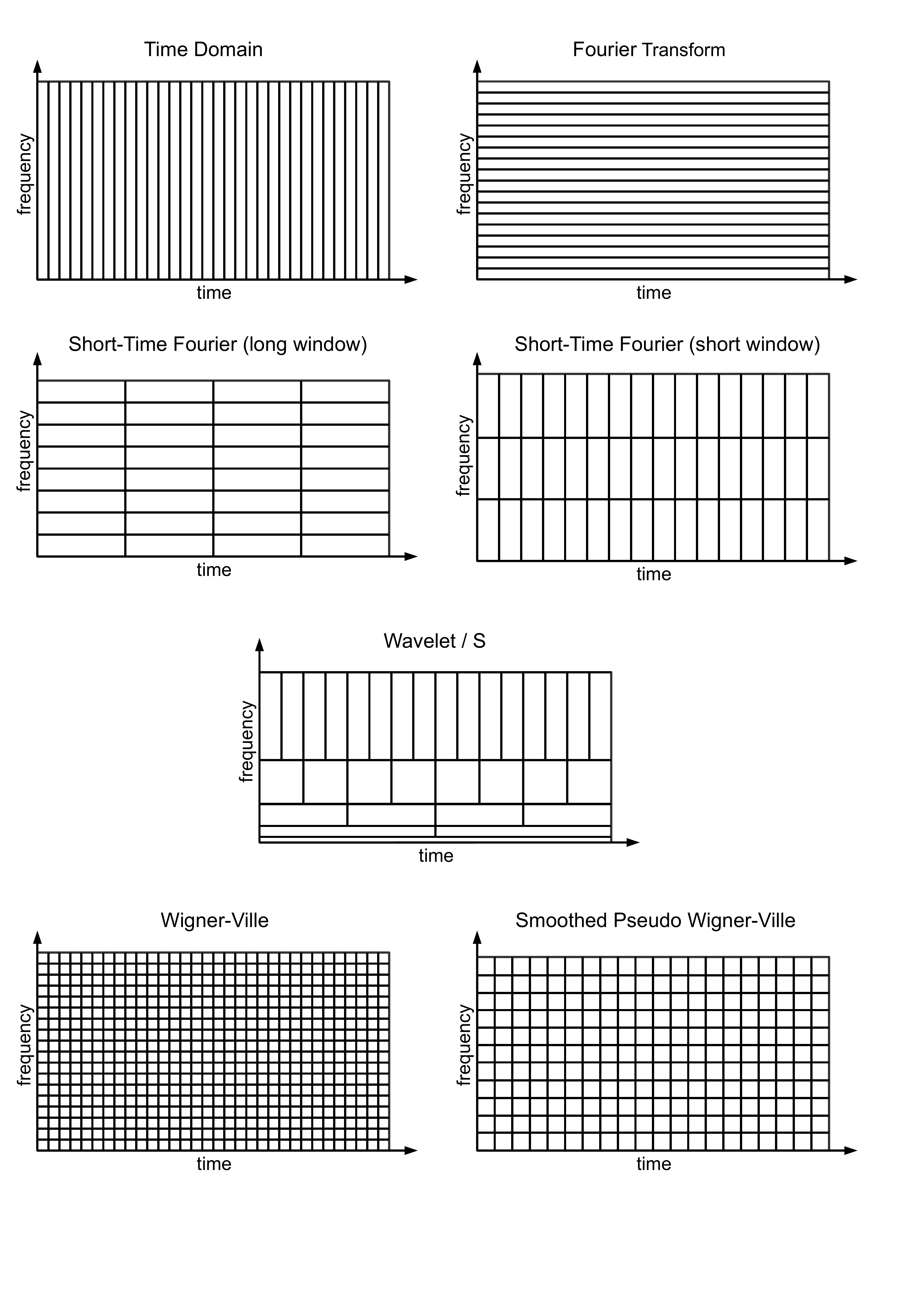}
\par\end{centering}

\centering{}\caption{\label{fig:comparison-tf-resolution}Schematics of the resolutions
of the transforms, including the time domain representation and the
standard Fourier spectrum for comparison}
\end{figure*}

The introduced time-frequency transforms are now compared with respect
to their resolution in time and frequency. Figure \ref{fig:comparison-tf-resolution}
presents an overview of the schematic resolution grids. It visualizes
the impact of the transforms and their parameters on the resolution.
For completeness, the figure also shows the plain time domain and
spectrum representations, that do not resolve frequency and time,
respectively.

\begin{sidewaysfigure*}
\begin{centering}
\includegraphics[scale=0.93]{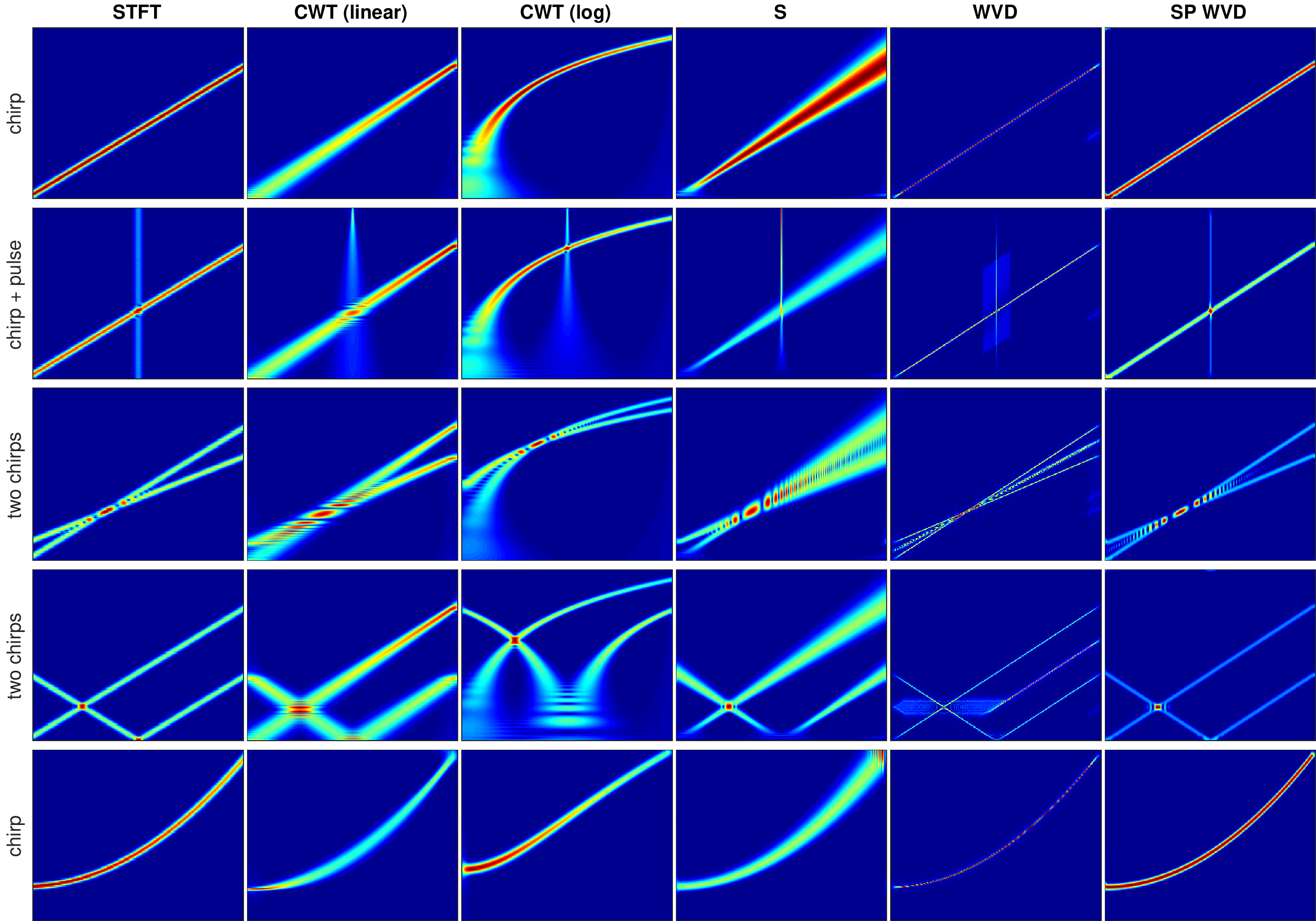}
\par\end{centering}

\caption{\label{fig:Synthetic-signals-(1)}Synthetic signals (1)}

\end{sidewaysfigure*}

\begin{sidewaysfigure*}
\begin{centering}
\includegraphics[scale=0.93]{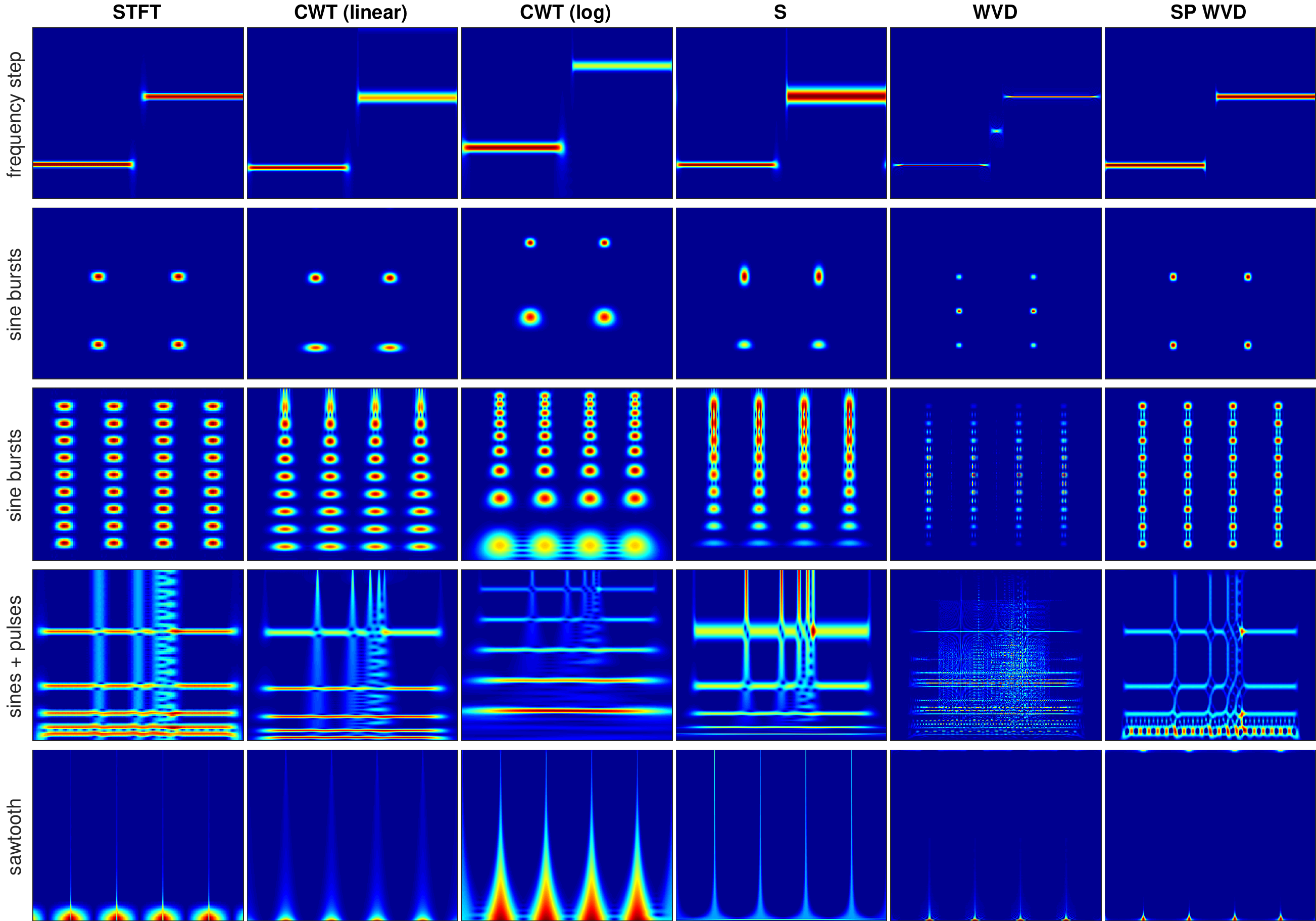}
\par\end{centering}

\caption{\label{fig:Synthetic-signals-(2)}Synthetic signals (2)}

\end{sidewaysfigure*}

\begin{sidewaysfigure*}
\begin{centering}
\includegraphics[scale=0.93]{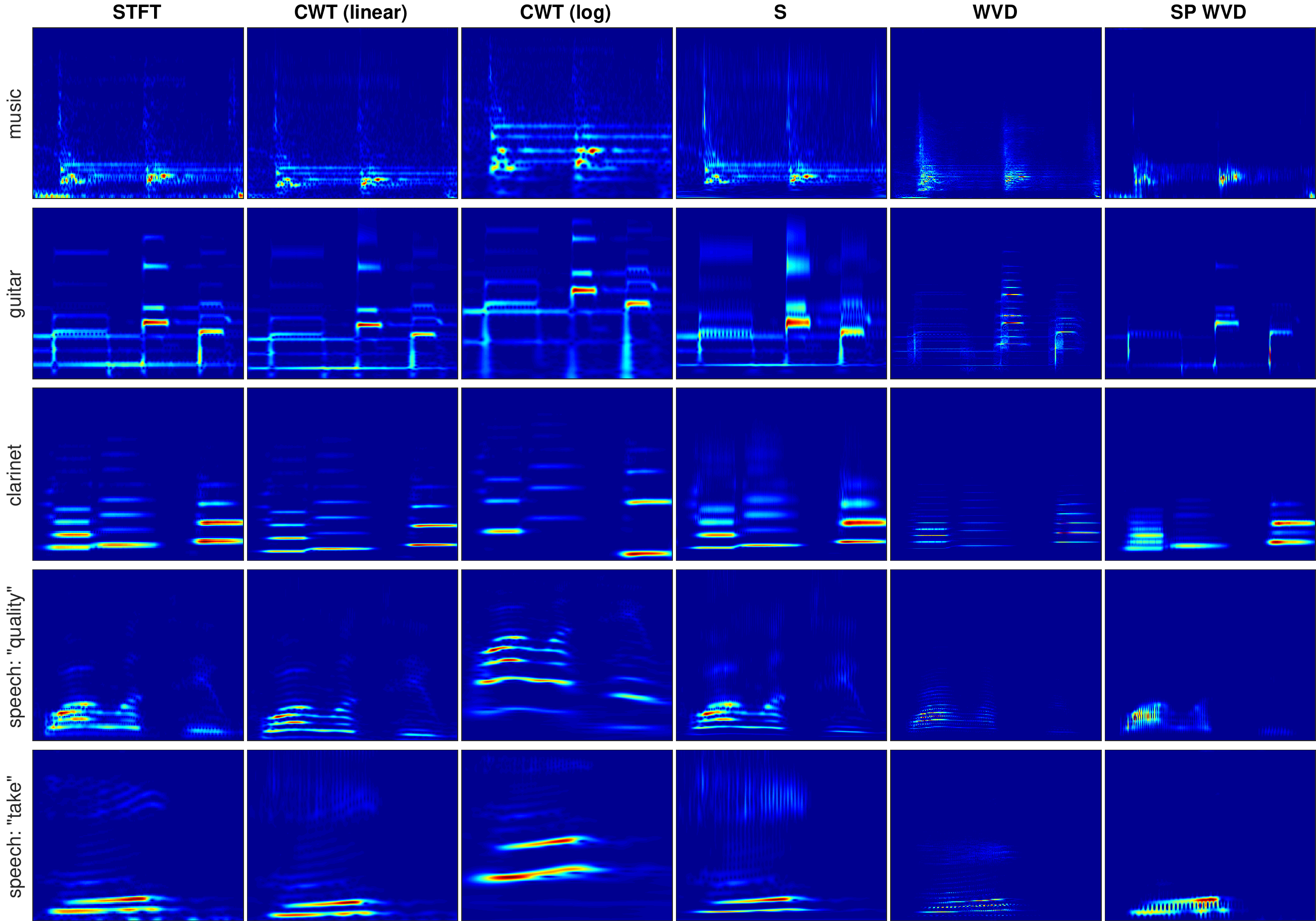}
\par\end{centering}

\caption{\label{fig:Audio-signals}Audio signals}
\end{sidewaysfigure*}

\begin{sidewaysfigure*}
\begin{centering}
\includegraphics[scale=0.93]{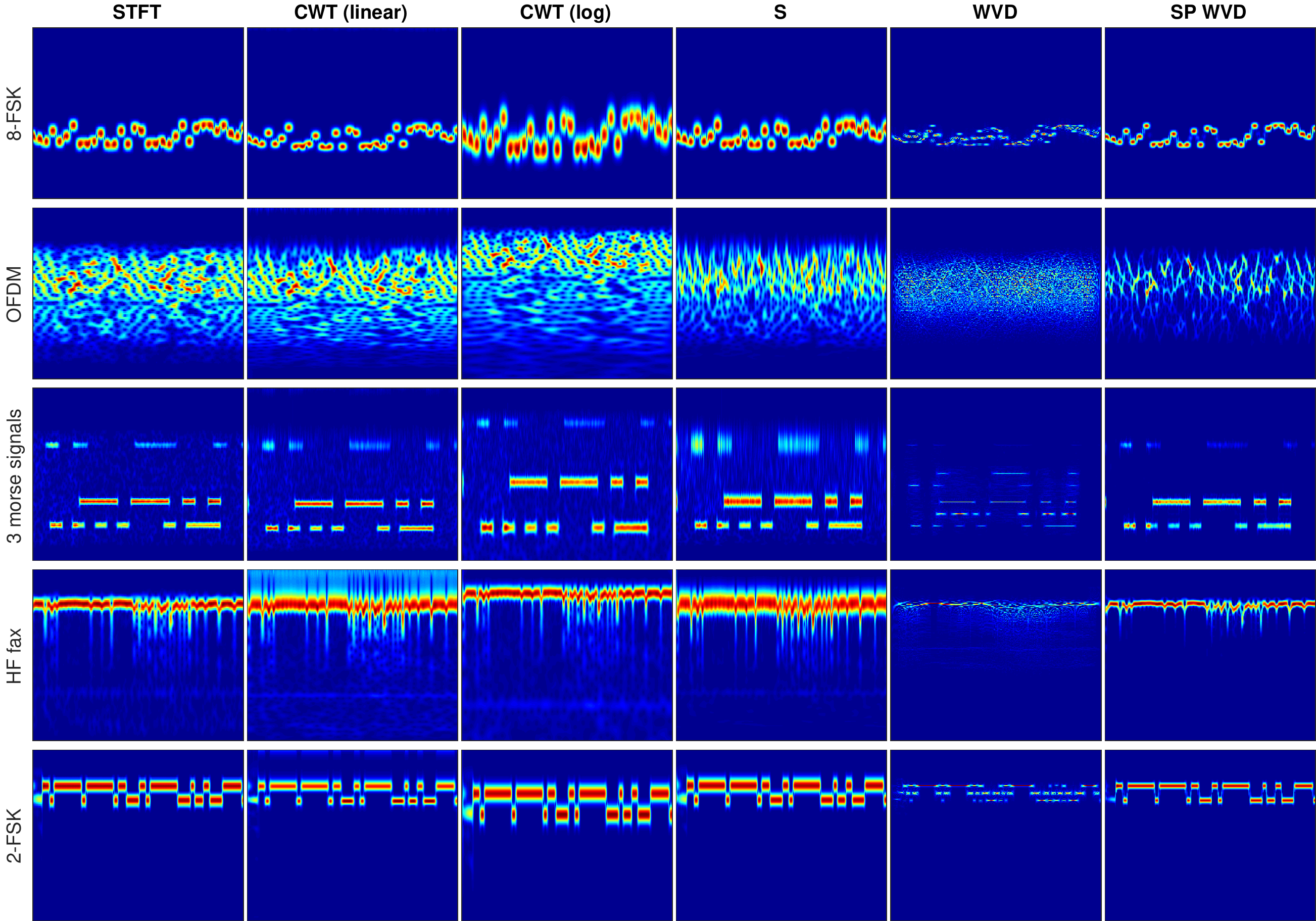}
\par\end{centering}

\caption{\label{fig:Radio-signals}Radio signals}
\end{sidewaysfigure*}

\begin{sidewaysfigure*}
\begin{centering}
\includegraphics[scale=0.93]{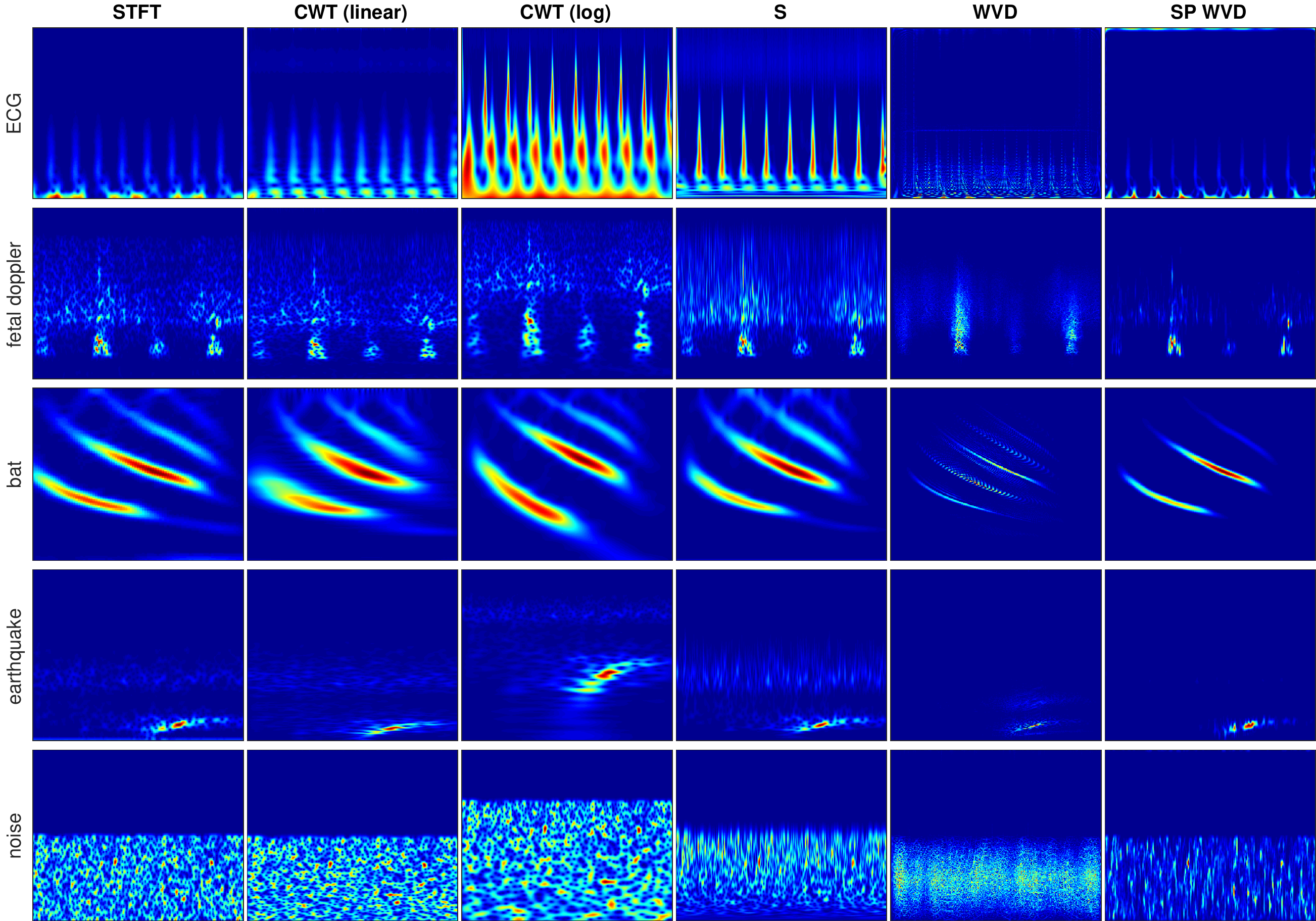}
\par\end{centering}

\caption{\label{fig:Signals-from-nature}Signals from nature and medical applications}
\end{sidewaysfigure*}

Furthermore, the transforms are compared for various synthetic signals
and signals from real sources that cover different applications. The
signals used for analysis consist of both synthetic and real-world
signals. Synthetic signals include sweeps, pulses, frequency steps
or short sine bursts. Some signals are composed of several of these
elements. The results for synthetic signals are shown in Fig. \ref{fig:Synthetic-signals-(1)}
and \ref{fig:Synthetic-signals-(2)}. Real-world signals include audio
signals from speech and music (Fig. \ref{fig:Audio-signals}), radio
signals with different modulation types (Fig. \ref{fig:Radio-signals})
and signals from nature and medical applications such as ECG, ultrasonic,
bat sound, earthquake (Fig. \ref{fig:Signals-from-nature}). 

The considered purpose of a transform is to visually present discriminative
time and frequency features. Therefore only the magnitudes of the
results are shown. Some transforms require certain parameters to be
tuned, e.g. window length of the STFT or the center frequency of the
Morlet mother wavelet. These parameters have been chosen such that
the features of the signals become easily visible. According to common
practice in literature, all transforms are typically plotted on linear
frequency scale, except for the wavelet transform, which occurs both
in logarithmic and linear scale.

Finally, the choice of the transform depends on the signal properties
and the further use or subsequent processing of the results. Important
aspects to consider are the desired resolution in time and frequency
as well as the tolerability of artifacts. Table \ref{tab:When-to-use}
summarizes the transforms and provides recommendations when to use
which transform.

\begin{table*}
\begin{centering}
\begin{tabular}{>{\raggedright}p{3.5cm}>{\raggedright}p{2.5cm}>{\raggedright}p{1.5cm}>{\raggedright}p{8cm}}
\toprule 
Transform & Time-Frequency Resolution & Artifacts & Application\tabularnewline
\midrule
Short-Time Fourier (STFT) / Gabor & poor & no & general purpose\tabularnewline
\addlinespace[0.1cm]
Continuous Wavelet (CWT) & poor,\\
frequency dependent & no & when variable time-frequency resolution required, e.g. audio\tabularnewline
\addlinespace[0.1cm]
Stockwell (S) & poor,\\
frequency dependent & no & when variable time-frequency resolution with a fixed phase alignment
required, emphasizes higher frequencies\tabularnewline
\addlinespace[0.1cm]
Smoothed Pseudo Wigner-Ville (SPWVD) & good & sometimes & general purpose, when high resolution required and some artifacts
can be tolerated\tabularnewline
\addlinespace[0.1cm]
Wigner-Ville (WVD) & excellent & strong & for simple signals or when artifacts can be tolerated\tabularnewline
\bottomrule
\addlinespace[0.1cm]
\end{tabular}\medskip{}

\par\end{centering}

\caption{\label{tab:When-to-use}When to use which transform}

\end{table*}

\bibliographystyle{IEEEtran}
\bibliography{tf_literatur}

\pagebreak{}

\end{document}